# Influence of grain size, shape and compaction on georadar waves: example of an Aeolian dune.

Julien Guillemoteau\*, Maksim Bano\*, Jean-Remi Dujardin\*

\* *Institut de Physique du Globe de Strasbourg & EOST, CNRS-UDS UMR 75-16, Strasbourg, France.*

Corresponding author: j.guillemoteau@unistra.fr

**SUMMARY**

Many Ground Penetrating Radar (GPR) profiles acquired in dry aeolian environment have shown good reflectivity inside present-day dunes. We show that the origin of this reflectivity is related to changes in grain size distribution, packing and/or grain shape in a sandy material. We integrate these three parameters into analytical models for bulk permittivity in order to predict the reflections and the velocity of GPR waves. We consider two GPR cross-sections acquired over Aeolian dunes in the Chadian desert. The 2D migration of GPR data suggests that dunes contain different kinds of bounding surfaces. We discuss and model three kinds of reflections using reasonable geological hypothesis about Aeolian sedimentation processes. The propagation and the reflection of radar waves are calculated using the 1D wavelet modelling method in spectral domain. The results of the forward modelling are in good accordance with real observed data.

**INTRODUCTION**

Ground Penetrating Radar (GPR) is a geophysical method based on electromagnetic wave (EM) propagation which is sensitive to dielectric permittivity contrasts. The resolution of the GPR method depends on the velocity of EM waves and the frequency of the antennae used. Following the λ/4 criterion (Widess, 1973; Jol, 1995 and Zeng, 2009), it varies from 5 cm to 22 cm for frequencies of 200–500 MHz and velocities of 0.1-0.18 m/ns. At this scale of analysis, the dielectric constant of the effective medium is controlled by the volumetric fraction of each sub-material (rocks matrix, air and eventually water) that constitute rocks. Therefore, one can say that the GPR response for rocks depends principally on the nature of its matrix, its porosity and its fluid saturation.

Many studies have shown that the GPR method can provide very detailed and continuous imaging of the internal structures of aeolian dunes and allow defining the relative chronology of sand deposition (Shenk *et al*, 1993; Bristow *et al*, 1996, 2005; Harari, 1996; Bano and Girard, 2001; Bristow *et al*, 2000; Neal and Roberts, 2001; Adetunji *et al*, 2008). Other studies considering the physical properties of rocks have shown that the properties that might change are the grain size distribution (Otto, 1938; Bagnold, 1941; Barndorff-Neilsen *et al*, 1982; Watson, 1986; Thomas, 1988; Lancaster, 1989, Wang et al., 2003), wind compaction (Bayard, 1947; Hunter, 1977) and/or the grain shape (Sen *et al*, 1981) in the case of sand/sandstone contact. These three structural parameters cause a macroscopic change of permittivity. Consequently, they should influence the GPR reflectivity within dry sand.

In this paper, we propose to explain these reflections by using a model of dry materials in which we can control the size, compaction and shape of the grain. First, we show how to incorporate granulometry data in the effective permittivity formulas for a mix of several grain sizes. Then, we propose a method to predict simultaneously both the porosity and permittivity of a bimodal mix of sand. For the third structural parameter, the grain shape, we simulate sand/sandstone contact by using the formula introduced by Sen *et al* (1981). Finally, we compute the propagation and reflection of electromagnetic waves within a two-layer model that depends on these three parameters and compare the modelled GPR response with real data acquired over two Aeolian sand dunes in the Chadian desert. We acknowledge that many other factors such as antenna radiation, ground coupling and scattering loss are not accounted for. In our analyses**,** these factors are assumed to change much more slowly than the effective permittivity in the case where reflections are observed in the data.

**INFLUENCE OF GRAIN SIZE DISTRIBUTION ON THE PERMITTIVITY**

The permittivity of rocks varies over a wide range of length scale. The GPR method is sensitive to permittivity contrasts for layers thicker than a few centimetres. However, the permittivity of each layer is controlled by smaller changes of permittivity occurring at the pore scale and/or by the ones caused by thinner laminations (around 1cm of thickness) which might be due to the daily alternating wind power**.** At this scale of analysis, the quasi-static approximation is valid and one can use the effective permittivity model for multiphase mixtures (Sihvola and Kong, 1988). In the case of sand dunes, the thin laminations are characterized by changes of granulometry. In the following, we attempt to compute the effective permittivity that depends on the volumetric fraction of each sub-material in the mixture. For example, a dry sand is a mix of air and quartz grains with relative permittivities of $\kappa_0 = 1$ and $\kappa_q = 5$, respectively. If one considers a material containing spherical grains of homogeneous size (only one grain size), the more appropriate model is the Maxwell-Garnett (MG) formula (Maxwell-Garnett, 1904) which is given by:

$$\kappa^* = \kappa_0 + 3(1-\phi)\kappa_0 \left[ \frac{\kappa_q - \kappa_0}{\kappa_q + 2\kappa_0 - (1-\phi)(\kappa_q - \kappa_0)} \right]. \quad (1)$$

On the other hand, if the material is constituted by grains with an infinite number of sizes, that is to say a non homogeneous grain size distribution, one can use the model described by the Hanai-Bruggeman-Sen (HBS) formula given for a mix of spherical grains and air (Hanai, 1968; Bruggeman, 1935; Sen et al., 1981) :

$$\left[ \frac{\kappa_q - \kappa^*}{\kappa_q - \kappa_0} \right] \left[ \frac{\kappa_0}{\kappa^*} \right]^{1/3} = \phi. \quad (2)$$

In these formulas, the parameter $\phi$ is the porosity of the rocks which characterises the volumetric fraction of each component. In reality, however, rocks exhibit neither a single nor an infinite number of grain size. Therefore, these two models should form the limits which might border the observed data. In order to predict the permittivity of materials with a finite number of sizes, one can use the Robinson-Friedman (RF) recurrence formula given for a mix of several grain sizes (Robinson and Friedman, 2001):

$$\kappa_n = \kappa_{n-1} + 3\kappa_{n-1} \frac{f_n}{\phi + \sum_{n=1}^{n} f_n} \cdot \frac{(\kappa_q - \kappa_{n-1})}{\kappa_q + 2\kappa_{n-1} - \frac{f_n}{\phi + \sum_{n=1}^{n} f_n}(\kappa_q - \kappa_{n-1})}. \quad (3)$$

The number $f_n$ is the volumetric fraction of the grains of size 'n' and has to validate the condition:

$$\sum_{n=1}^{N} f_n = 1 - \phi, \quad (4)$$

where N is the total number of sizes. Equation (3) might be interpreted as an implementation of the MG model. It has already been studied experimentally with several mixes of glass beads (Robinson and Friedman, 2001, 2005).

Usually, the grain size distribution is measured in terms of percentage of weight. In our study, we assume that the sand is composed exclusively of quartz of the same density. This allows us to consider the percentage of weight as being equivalent to the percentage of solid volume. In equation (3), the volumetric fraction $f_n$ of solid is related to the grain size distribution $A_n$ and the porosity $\phi$ by the following equation:

$$f_n = (1-\phi)\frac{A_n}{\sum_1^N A_n}, \qquad (5)$$

In Figure 1, we show the implementation of equation (3) for three different synthetic granulometry (grain size distribution $A_n$) data indicated by D1, D2 and D3, respectively. We take a range of grain sizes between 0.04 and 1mm. As expected, in the case of a homogeneous distribution (D1: only one size of 0.48 mm) the result (blue line) fits with the MG model (equation 1). If we consider a non-homogeneous distribution (D3: 25 sizes in the range of 0.04-1mm), the results (green line) tends toward the HBS model (equation 2), while in the intermediate case (D2: four sizes of 0.24, 0.48, 0.6 and 0.72 mm, respectively) the result (red line) is bounded by the two previous cases. These observations are in agreement with the fact that dry sand permittivity should be located between the two extreme cases which are described by the MG and HBS models, respectively.

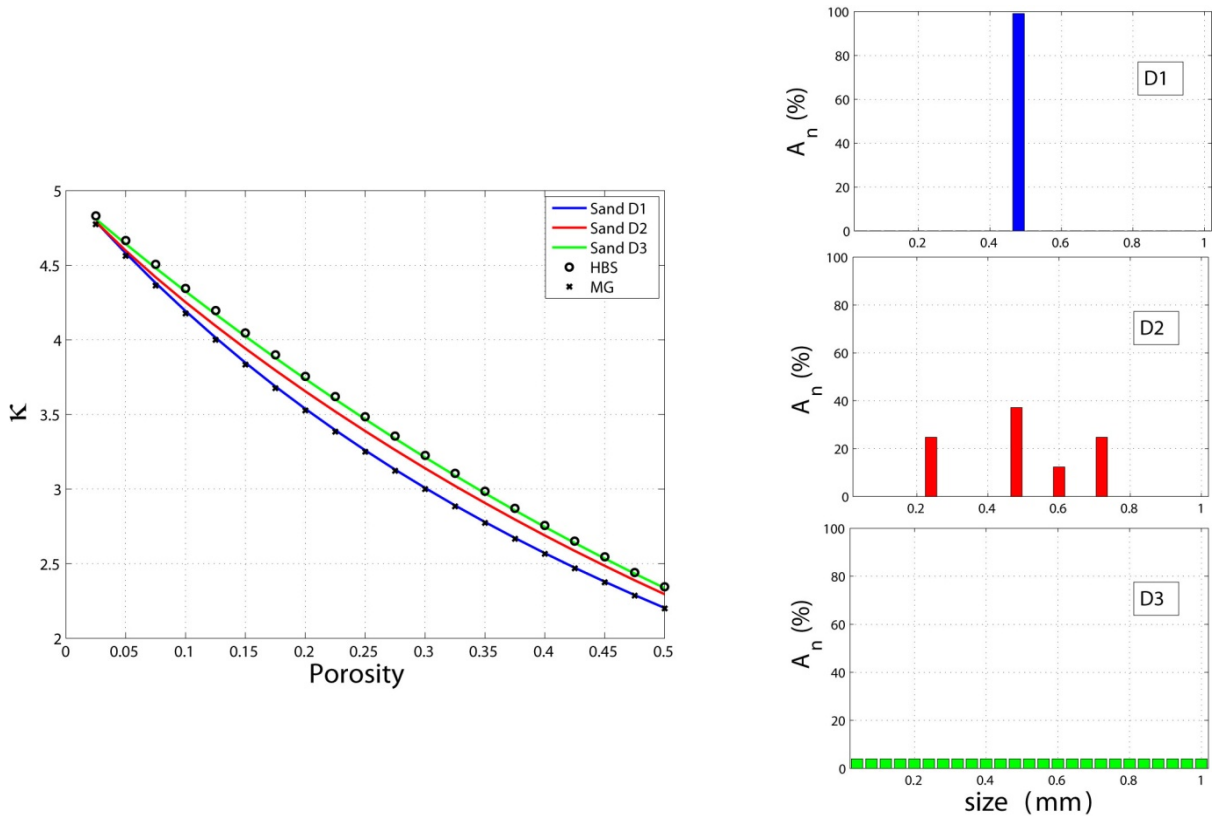

**Figure 1.** *Influence of the granulometry (grain size distribution $A_n$) on permittivity model. The permittivity is plotted for a mix of quartz ($\kappa_q = 5$) and air ($\kappa_0 = 1$). D1: one size of 0.48 mm. D2: four sizes of 0.24, 0.48, 0.6 and 0.72 mm, respectively D3: 25 sizes in the range of 0.04-1mm)*

For a constant porosity, the RF model (equation 3) predicts a change of permittivity caused by the granulometry. Although this change is quite small, it might cause significant changes to the reflectivity of radar waves if we consider materials having small permittivity values, as is the case in dry sand dunes. However, some studies have shown that the grain size distribution controls the porosity as well. Therefore one has to bear in mind that the two parameters $f_n$ and $\phi$ are joined (interfere with each other).

**INFLUENCE OF THE GRANULOMETRY ON THE POROSITY**

It is very difficult to use the granulometry to predict the porosity of a mix with more than two sizes. Indeed, while there is an analytical model that considers a bimodal mix of spheres (McGeary, 1961; Marion, 1992), it cannot be accurately adapted to rocks because the shape of their components is not totally spherical. In this part, we present the 'fractional packing model' introduced by Koltermann and Gorelick, (1995). This model is a semi-empirical law that gives the porosity of a bimodal mix. Let us consider a bimodal material composed of two parts: a 'small grain' and a 'large grain' that are characterised by their porosity $\phi_{small}$ and $\phi_{large}$, respectively. Each phase has a volumetric fraction $F$ that satisfies the following condition:

$$F_{small} + F_{large} = 1 \quad . \tag{6}$$

It is important to note that $F$ is different from the volumetric fraction of solid $f_n$ introduced in the RF model. Indeed, the sum of the $f_n$ has to fulfil the condition:

$$f_{small} + f_{large} = 1 - \phi_{mix} \quad , \tag{7}$$

where $\phi_{mix}$ is the porosity of the mixture. The volumetric fraction of the solid $f_n$ and the one of the phase $F$ are related for each part by the equation:

$$f_{small} = (1 - \phi_{small})F_{small} \tag{8}$$

The same relation is valid for the 'large grain' part. The 'fractional packing model' is separated in two regimes which correspond to different packing properties: i) the coarse packing where fine grains are disposed inside the pore space of coarse grains (first part of the curve in Figure 2), and ii) the fine packing (second part of the curve in Figure 2) where coarse grains are disposed inside a fine grain matrix. All of these cases are described in Figure 2. This model depends on an empirical parameter which is related to the ratio between the two sizes. In this paper we present only the resulting porosity for several ratios between two grain sizes ($R_{large}/R_{small}$ = 5; 10 and 20, respectively). In Figure 2, the fractional packing model is plotted for different ratio of sizes using a porosity of 0.48 which is in good agreement with values found in the literature for sand dunes (Atkins and McBride, 1992). One can see that the bigger the size ratio ($R_{large}/R_{small}$), the lower the corresponding porosity value. The minimum porosity occurs when the volumetric fraction of the 'small grain' part is equal to the porosity of the 'large grain' part. This case corresponds to the central part of Figure 2.

So far we have seen that the permittivity depends directly on the granulometry, as it is shown in the previous section. On the other hand, the granulometry indirectly influences the permittivity by controlling the porosity, as discussed in the present section. Therefore, in order to predict the permittivity of dry sand, we have to take into account not only the size distribution but also the porosity. Both of these parameters are related and are controlled by each other. In our study, we model the permittivity of effective bimodal sand using equation (3) where the porosity is estimated by the fractional packing model.

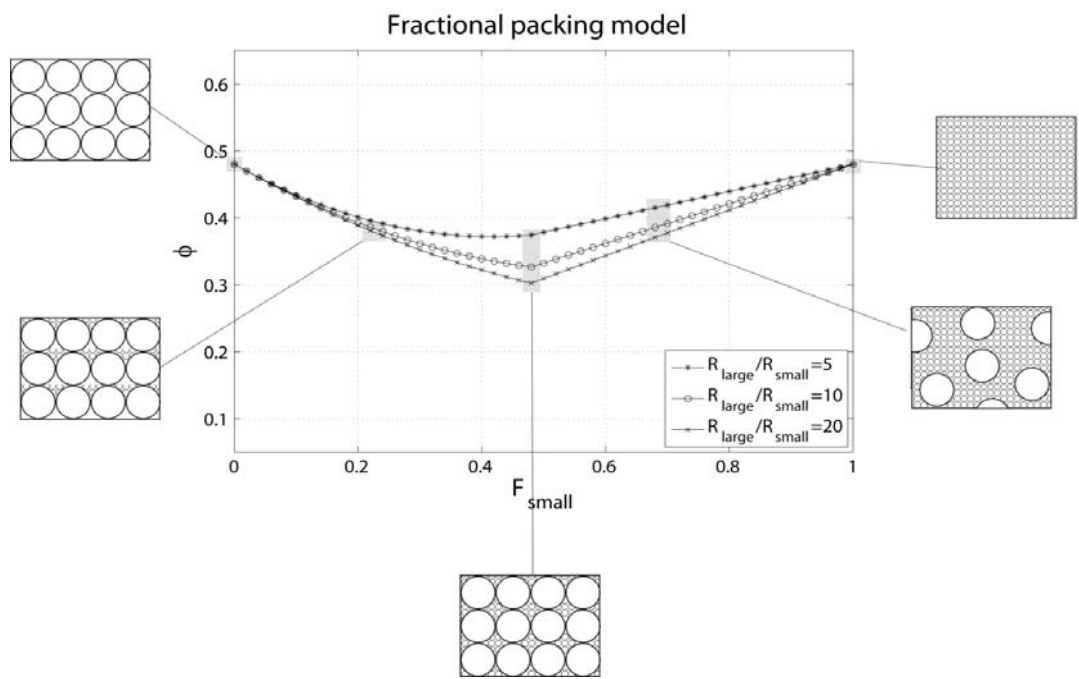

**Figure 2.** *Fractional packing model introduced by Koltermann and Gorelick, (1995). The porosity of each phase is taken to be equal to 0,48. The minimum of porosity corresponds to the case $F_{small} = \phi_{large}$.*

Figure 3 shows the GPR reflection coefficient for four different size ratios ($R_{large}/R_{small}$ = 3; 5; 10 and 20, respectively). The results of Figure 3 are computed with equation (12) for a contact between two bimodal sands with volumetric fraction of the "small grain" part $F^1_{small}$ and $F^2_{small}$ for a normally incident wave. The results indicate that, in some cases, the reflectivity can reach 8%. One will notice that the higher the size ratio, the stronger the reflection coefficient. However, this only occurs where there is a strong change of grain size distribution which is the case of contact between well sorted sand (only one grain size: where the part F of one of the two sizes is equal to 0 or 1) and bimodal sand ($F_{small} \approx \phi_{large}$).

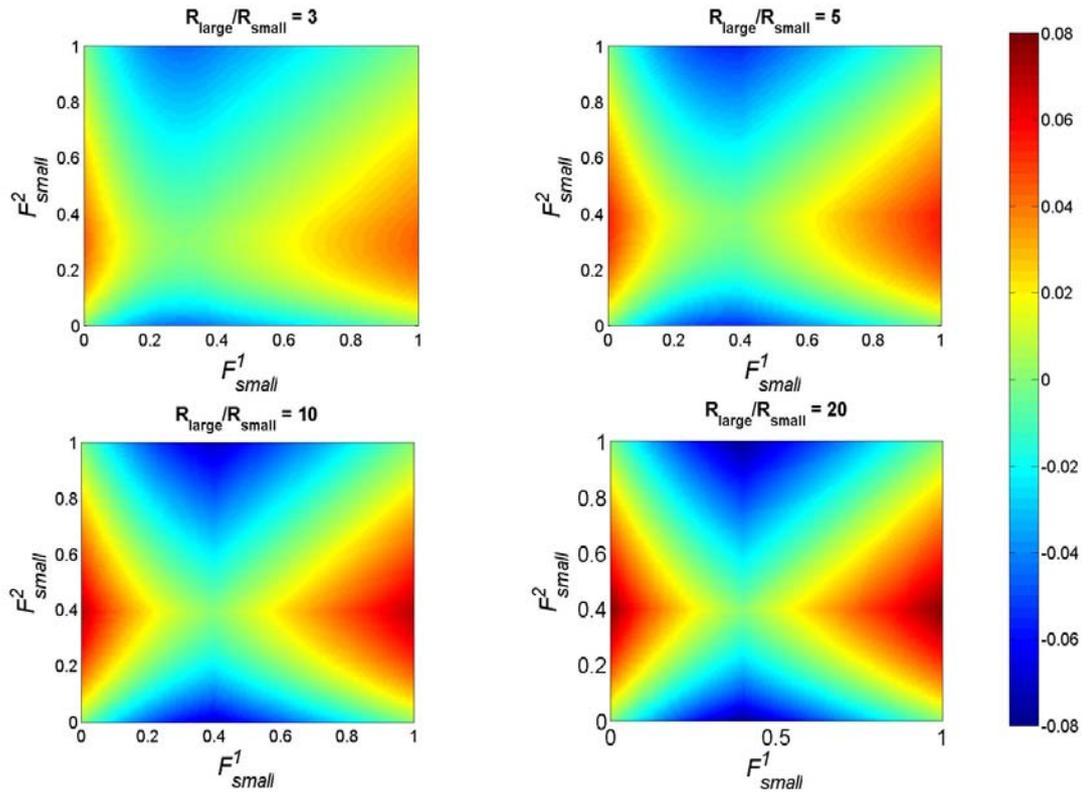

**Figure 3.** *Theoretical reflectivity between two sandy horizontal layers with different grain size distribution. The strongest reflectivity occurs where there is contact between a well sorted sand ( $F_{small} \approx 0$ or 1) and a bimodal mix ( $F_{small} \approx \phi_{large}$ ).*

**INFLUENCE OF GRAIN SHAPE ON THE PERMITTIVITY**

Changes in grain shape influence the level of connectivity of the material and influence the macroscopic electrical properties (Jackson et al, 1978; Sen et al, 1981). The equations (1) and (2) are only valid for spherical grain. However, these laws can be generalized for all elliptic shapes by introducing a factor L depending on the shape of the grains. In this context, the MG formula can be written as follows (Stratton, 1941):

$$\kappa^* = \kappa_0 \left[ \frac{\kappa_0 + (\kappa_q - \kappa_0)(1 - \phi + \phi L)}{\kappa_0 + L\phi(\kappa_q - \kappa_0)} \right], \qquad (9)$$

and the HBS model as follows (Sen et al, 1981; Greaves et al, 1996):

$$\left[ \frac{\kappa_q - \kappa^*}{\kappa_q - \kappa_0} \right] \left[ \frac{\kappa_0}{\kappa^*} \right]^L = \phi . \qquad (10)$$

The parameter L is called the depolarization factor. It depends on the three elliptical radiuses (Mendelson and Cohen, 1982; Boyle, 1985; Asami, 2002). Non spherical grain yields to macroscopic anisotropy.

| Shape | Name | Axes | Depolarising factor |
|---|---|---|---|
| 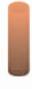 | Cylinder | $a = b \ll c$ | $L=0$ |
| 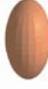 | Sphere 'Prolate' | $a = b < c$ | $0 < L < 1/3$ |
| 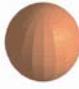 | Sphere | $a = b = c$ | $L = 1/3$ |
| 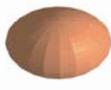 | Sphere 'Oblate' | $a = b > c$ | $1/3 < L < 1$ |
| 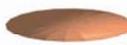 | Disk | $a = b \gg c$ | $L = 1$ |

**Table 1:** *Depolarising factor for different grain shape.*

Theoretically, L is a vector composed of three components with values between 0 and 1; the sum of these values is equal to unity. Each component corresponds to one direction of anisotropy. In our study, since we deal with 1D modeling for rays in the quasi-vertical direction (see below), we take L to be the scalar equivalent of the z-component of the anisotropic depolarizing factor. The values of L for different kinds of grain shape are presented in Table 1. By using these characteristics, one can model sand by mixing a set of spherical grains (L=1/3) and a sandstone by mixing only a set of oblate spheres (L=0.5) as it is suggested in Jackson et al (1978) and Greaves et al, (1996). ). Introducing L = 1/3 in equation (9) one can easily find the MG formula given in (1) for a well sorted sand (containing identical spherical grains).

**1D WAVELET MODELLING AND COMPARISON WITH REAL GPR DATA**

The propagation and the reflection of the radar wave are calculated using the 1D wavelet modelling in the Fourier domain. This method, discussed by Bano (1996, 2004), gives the complex spectrum of a normally incident wavelet travelling through a homogeneous absorbing layer from *z* = 0 and reflected at a depth *z* as follows:

$$\vec{E}(z,\omega) = G(z) R_{12} \vec{E}(0,\omega) e^{i\omega\tau} e^{-\alpha 2z}, \qquad (11)$$

Where $\tau = 2z/V$ is the two way propagation time, V is the propagation velocity of EM waves, $R_{12}$ is the reflection coefficient, $G(z)$ accounts for the geometrical spreading while $\vec{E}(0,\omega)$ is the complex spectrum of the electrical source of radar data at z = 0. The first exponential function models the propagation, and the second one considers the amplitude term related to the attenuation properties of the media. In our case, we consider a non-dispersive and non-attenuating medium ($\alpha = 0$). This is a good assumption for a material composed of dry sand, as it is the case of the dry dunes in the Chadian desert (see below).

The reflection coefficient $R_{12}$ of a wave propagating through a homogeneous media with permittivity $\kappa_1$ and normally reflected at the interface of a second layer with the permittivity $\kappa_2$ is given by the following formula:

$$R_{12} = \frac{\sqrt{\kappa_1} - \sqrt{\kappa_2}}{\sqrt{\kappa_1} + \sqrt{\kappa_2}}. \qquad (12)$$

The GPR data shown in Fig. 4a were collected in a dry sand dune in the Chadian desert using shielded antennas of 450 MHz. The acquisition was on transverse electrical (TE) mode with constant offset of 0.25 m and the antennas were moved by steps of 0.125 m. The first events in Fig. 4 are the direct air waves followed immediately by the direct ground waves; these events are superimposed. In this study, we use the first event of each trace as a source wavelet for the modelling of the GPR reflections inside the dune. This choice is largely discussed and justified in Bano (2004). The undulating reflection indicated by white arrows in Figure 4a shows the base of the dune (which in fact is flat, see migration section in Figure 4b and consists of pebbles (over 2.0 mm in diameter). This reflection is from the contact between the aeolian sands and the deeper lake deposits that consist of an unconsolidated silty sandstone layer of very fine to medium grain size. In Figure 4a, we observe a nice diffraction hyperbola situated just under the base of the dune (see green circle), which fits very well with a velocity of 0.18 m/ns. This value represents a sort of average velocity from the top of the dune to the diffraction point and is in good agreement with values found in literature for dry sand (Von Hippel, 1954; Costas et al., 2006; Gómez-Ortiz et al., 2009). The wavelength for the dominant frequency of 450 MHz and velocity V=0.18 m/ns is 40 cm ($\lambda/4 = 10$ cm). This value is much larger than the grain size (from 0.02-2 mm diameter) and thin laminations inside the dunes.

In Figure 4b we present the topographic Kirchhoff migration as it is discussed in Lehmann and Green (2000). The velocity used is constant and equal to 0.18 m/ns corresponding to an average permittivity of 2.8 for dry sand. A schematic interpretation of the migrated section is shown in Figure 4c. The dune exhibits a complex internal structure with numerous bounding surfaces and it is composed of a minimum of four smaller dunes (indicated by 1, 2, 3 and 4, respectively, in Figure 4c). From this Figure, we can identify three kinds of reflectors:

- The ones which show a dip angle inside these small dunes (blue lines).
- The ones which may describe the surface of the internal small dunes (yellow lines).

-The reflector at the base of the system which is related to the contact between the sand and the bedrocks (brown line).

The first family of reflectors takes place within small dunes and is characterized by a large dip angle (between 15 and 25°). Some of these reflectors are shown in blue in Figure 4c. Otto (1938) and Bagnold (1941) suggested that these internal laminations are probably related to the fluctuation of transporting power which may segregate grain size differently over time. Moreover, these aeolian deposits may alternate with avalanching (Hunter, 1977, Mountney and Howell, 2000).

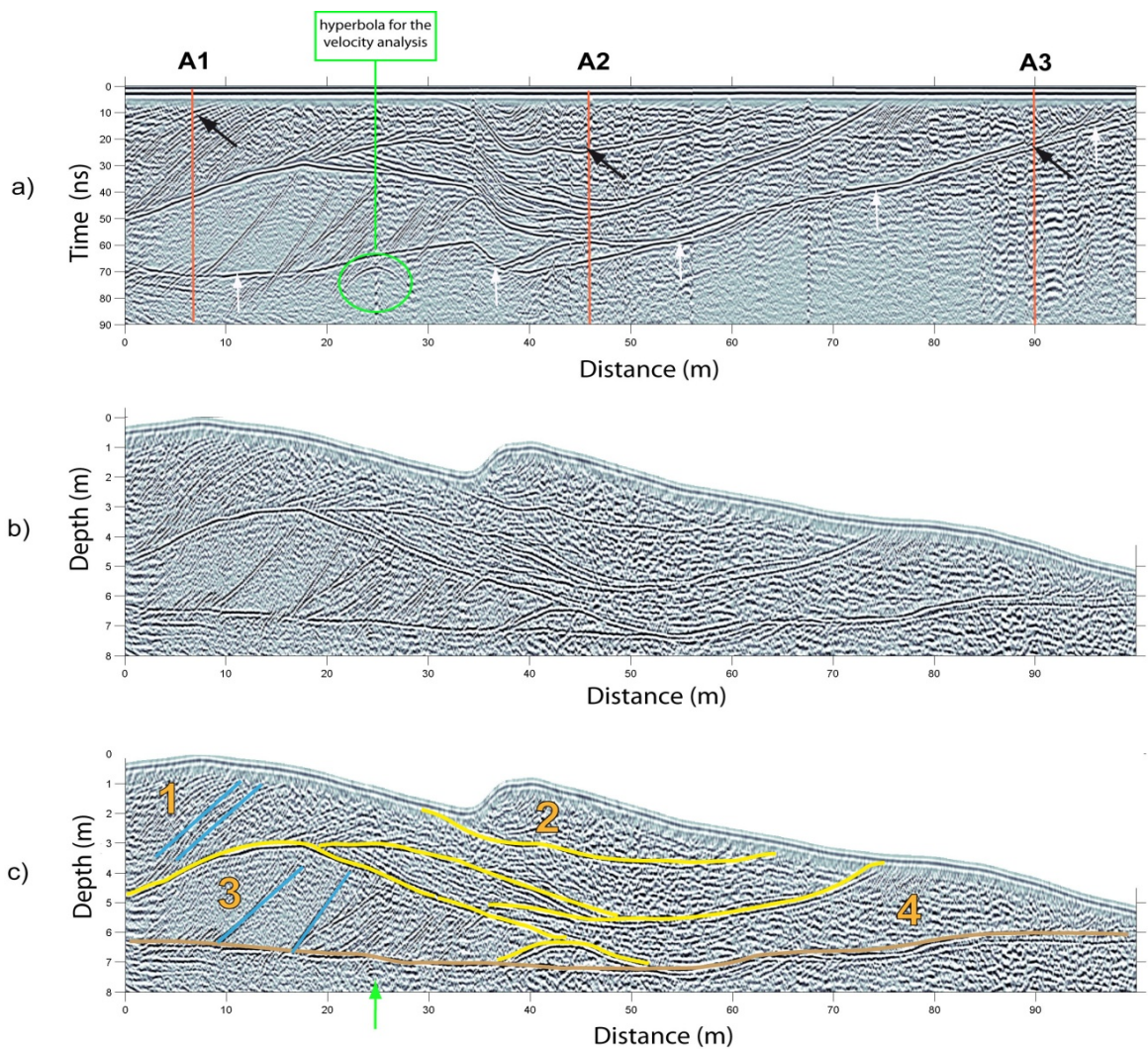

*Figure 4.* *a) The GPR section collected in a dry sand dune in the Chadian desert using shielded antennas of 450 MHz. The three vertical red lines refer to the traces which are modelled in this paper. b) The topographic Kirchhoff migration using a constant velocity of 0.18 m/ns. c) Interpretation considering three kinds of bounding surface. Blue: change in granulometry. Yellow: change in compaction. Brown: contact sand /sandstone.*

If we consider the second family of reflectors (related to the old surfaces of dunes) and according to the observations of Bayard (1947) in Mauritania, the sand at the surface of dunes is compacted by the wind over few centimeters. Hunter (1977) defined these reflectors as the "wind ripple surface", while Mountney and Howell (2000) observed a bimodal grain size distribution. As a consequence, the packing properties (for example the porosity $\phi$) of this sand should differ from the sand inside the dune which may result, at least partly, from wind deposition and avalanching of the leeward slope.

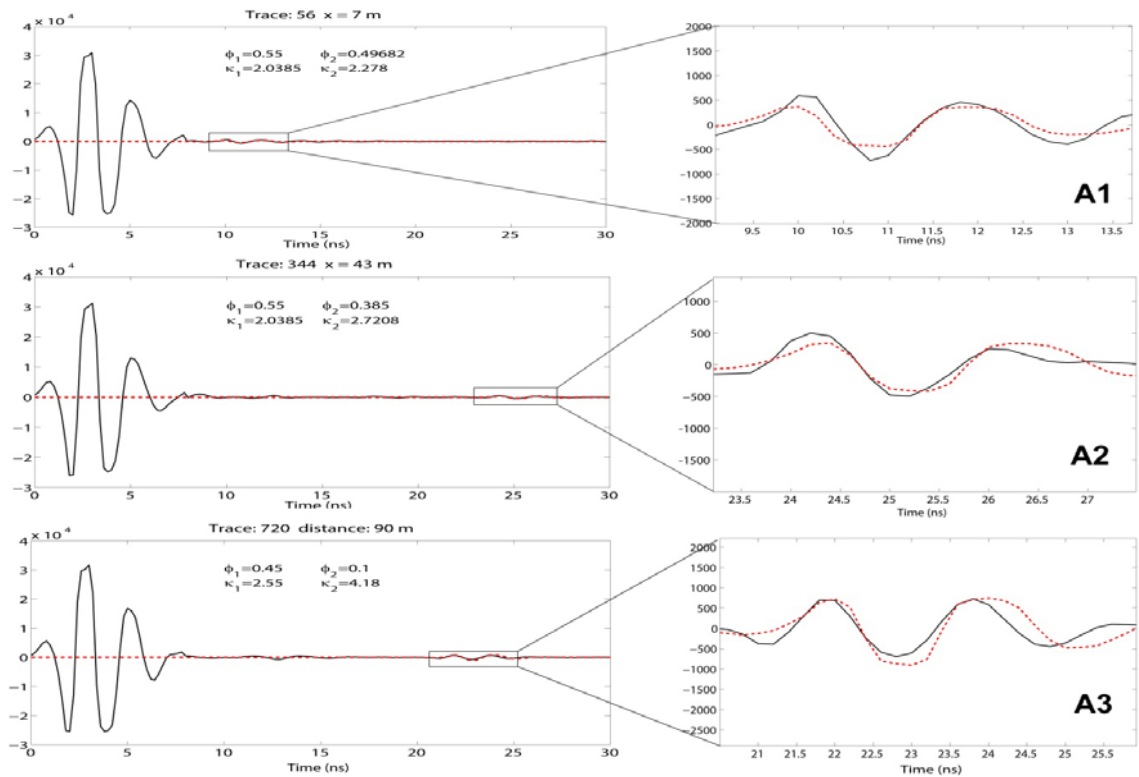

*Figure 5. Comparison between modelled (in red) and observed (in black) GPR data for the three kinds of physical change presented in this study: size distribution (top), compaction and size distribution (middle) and sand/sandstone contact (bottom). The wavelets presented are located at x=12 m, x=43 m and x=90 m on the section shown in Figure 4a.*

Finally, the third kind of reflector is the bottom of the dune, which indicates the contact between aeolian sand and sandstone (see the brown line of Figure 4c). The reflection coming from this latter reflector seems to be relatively strong (see the white arrows in Figure 4a), which justifies our choice of a non-absorbing media and the consequential low probability of having water saturation inside the dune.

In the following analysis we model the radar reflections generated by these three reflectors. On the other hand, owing to the high velocity of the sand (V = 0.18 m/ns) and according to Annan et al. (1975) and Jiao et al (2000), for an antenna located on the air–sand interface, the TE mode radiation has the largest amplitude at the critical angle $\theta_c = \sin^{-1}(V/0.3) = 37°$. Under this condition, the radiation pattern of the antenna is almost unchanged for angles between 0 and 25 degrees. Since we deal with 1D modeling for layers having a dip-angle between 0° and 25° (or for rays which deviate from the vertical on the range of 0-25°), the radiation pattern variation is not accounted for in equation (11).

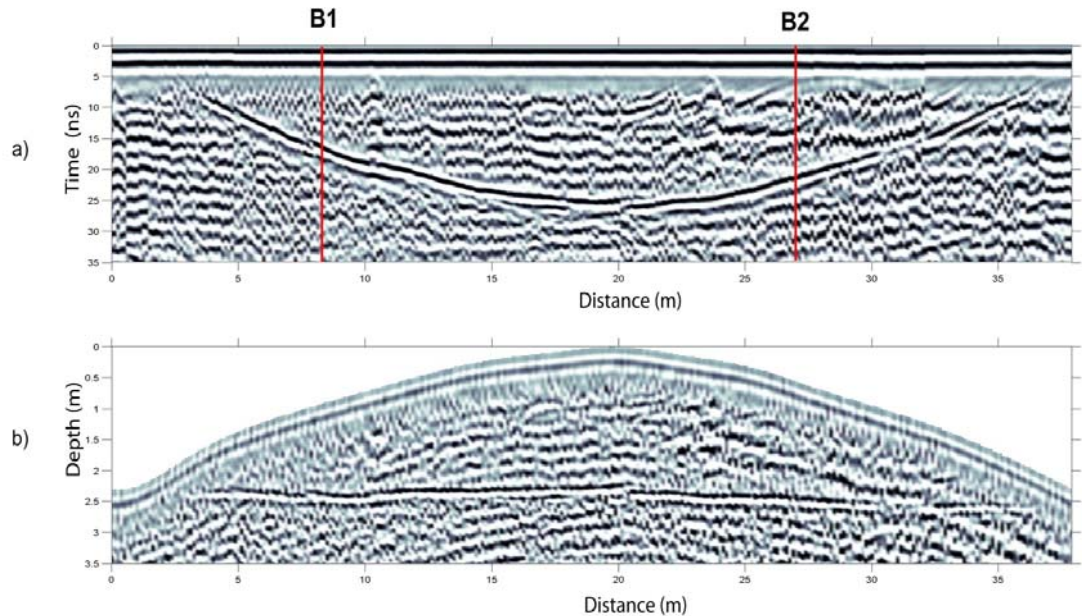

*Figure 6. a) The GPR section collected in a dry small dune using shielded antennas of 450 MHz. b) The topographic Kirchhoff migration using a constant velocity of 0.18 m/ns. The two vertical red lines refer to the traces which are modelled in this paper.*

Thus, the spectrum of a wavelet reflected from a reflector inside the dune is calculated by using equation (11), with the reflection coefficient given by (12). The permittivities of the two layers (in equation 12) are estimated by using the equations (3) and (5), while the porosity $\phi$ is predicted by using the empirical modelling of Koltermann and Gorelick (1995). Although the traveltime $\tau$ (in equation 11) can be read directly from the radar sections, we adjust it by minimizing the difference between the theoretical and real wavelets within a reasonable time interval. Afterwards, the theoretical reflected wavelet in time (obtained by transforming back the spectrum given in 11) is compared with the observed reflections.

The three observed reflections A1, A2 and A3 indicated by black arrows in Figure 4 are supposed to be related to the three families of reflectors previously discussed. The comparisons of synthetic reflected wavelets (in red) with three different real observed wavelets (A1, A2 and A3 in black) are shown in Figure 5. The observed wavelet A1 (top) is compared to the theoretical GPR response (in red) that is supposed to be reflected from contact between mono-modal sand and a mixing of two sands with size ratio of 2. Here we note that this structural change, which is characterized by a small contrast of permittivity (2/2.25), causes a significant reflectivity. The observed wavelet A2 (middle) is compared to the theoretical wavelet calculated for a contact between un-compacted well-sorted sand (one grain size) with porosity $\phi = 0.55$ and compacted bimodal sand with the total porosity $\phi = 0.39$. The comparison between the observed wavelet A3 and the synthetic wavelet is shown on the bottom of Figure 5. The permittivities are computed using the HBS formula (10). The synthetic wavelet is calculated for a contact between dry sand ($\phi = 0.45$ and $L = 1/3$) and sandstone which corresponds to the reflection from the base of the dune. A good fit is found by considering a sandstone composed of oblate grains ($L = 0.5$ in equation 10) with porosity $\phi = 0.1$. The effective permittivities ($\kappa_1$ and $\kappa_2$) of the two layers are shown within each graphic. They vary from 2-3 for the dune to 4.2 for the sandstone and are in good agreement with the values given by Davis and Annan (1989). It is interesting to notice that the value of permittivity for the dry sand estimated here is in good accordance with the value found by performing migration analysis of the diffraction hyperbola presented in Figure 4a. Remember that the best migration velocity is equal to 0.18 m/ns which cores+ponds to a permittivity of 2.8 for dry sand.

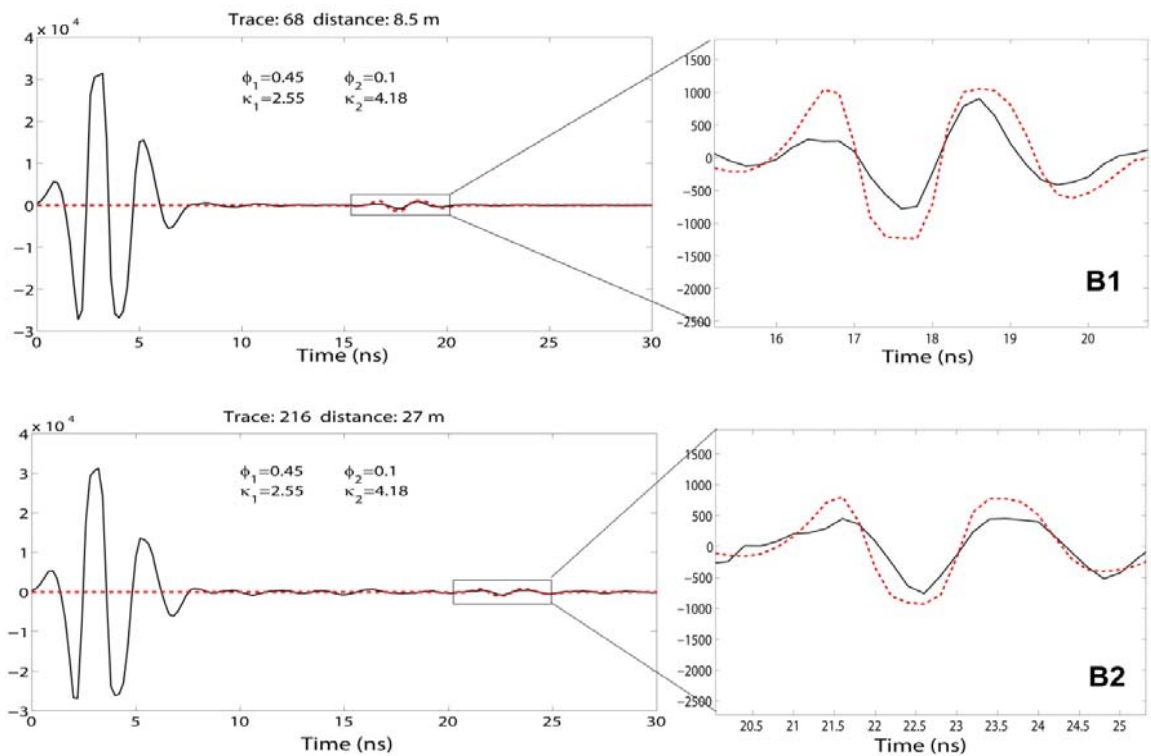

*Figure 7. Comparison between modelled (in red) and observed (in black) GPR data for sand/sandstone contact. The wavelets are located at x=8.5 m and x=27 m on the section shown in Figure 6.*

Figure 6a shows a second example of GPR data collected from a dry small dune using shielded antennas of 450 MHz. The bent reflection becomes a flat reflector after topographic migration with a velocity of 0.18 m/ns (see Figure 6b). This reflector indicates the contact between the dry sand and sandstone and corresponds to the base of the dune. To model the reflections from the bottom of the dune, the permittivities of the two layers are estimated by using equation (10). We take L = 0.33 for the first layer (dry sand composed of spherical grains) and L = 0.5 for the second layer (sandstone composed of oblate grains). Two observed wavelets (B1 and B2 located to the traces indicated by vertical red lines in Figure 6a) are compared with calculated wavelets and the results are shown in Figure 7. The porosity of the sand is taken equal to 0.45 and a reasonable fit between the modelled and measured wavelets is found for a porosity of 0.1 for the sandstone.

**CONCLUSION**

In this study, we show that changes in grain size distribution, packing or grain shape are able to cause contrast in dielectric permittivity that is detectable by GPR. In order to predict the effective permittivity of dry sand, we use three MG, HBS and RF relationships, respectively. Using reasonable geological hypothesis about the properties of sand, we can use GPR modelling to explain the different boundaries encountered within dunes. We assume that the inclined reflectors are due to changes of grain size distribution resulting from the variation of wind depositions. We incorporate granulometry data in the effective permittivity formulas for a mixture of perfect spheres. By this way, we model an average granulometry of the thin laminations which are not detectable by GPR reflectivity (due to high frequency alternating wind power). If we know precisely the permittivity of each small lamination (which depends on its granulometry), another way to estimate the effective permittivity, would have been to consider the effective medium as a mix of N thin plates (L=1 in equation 10) of sand with volumetric fraction of each lamination proportional to its thickness. On the other hand, we associate the big reflectors, which characterizes small internal dunes, to bimodal sand compacted by the wind ('wind ripple surface'). Finally, we show that the largest contrast of permittivity corresponds to the base of the dune. We model it as the contact between sand and sandstone by using the HBS formula.

The results of the modelling show good accordance with GPR data obtained over two different dunes in the Chadian desert. However, the information contained in GPR data does not allow inverting the parameters of sands directly (due to equivalent models) without additional a priori information. In this paper we only show that it is possible to explain the observed GPR by using a forward modelling that takes into account grain size distribution, packing and/or grain shape. Combining the application of the proposed method with additional constraining information from cores or trenches is the next logical step.

**ACKNOWLEDGMENTS**
The authors thank the Editor Mark Everett and an anonymous reviewer for their constructive comments and helpful suggestions which considerably improved the quality of the manuscript.

Published in Geophysical Journal International (2012), Vol. 190, p 1455-1463
doi: 10.1111/j.1365-246X.2012.05577.x...